\documentclass[a4paper,twocolumn,10pt,unpublished,unpublished]{quantumarticle}
\pdfoutput=1
\usepackage[utf8]{inputenc}
\usepackage[english]{babel}
\usepackage[T1]{fontenc}
\usepackage[numbers,sort&compress]{natbib}
\usepackage{placeins}
\usepackage{amsfonts}
\usepackage{pgfplots}
\usepackage{mathrsfs}
\usepackage{physics}
\usepackage[caption = false]{subfig}

\usepackage{tikz}
\usetikzlibrary{quantikz}
\usepackage{float}

\usepackage[colorlinks=true,citecolor=blue,linkcolor=magenta]{hyperref}

\renewcommand{\vec}[1]{\boldsymbol{#1}}  

\pgfplotsset{compat=1.16}
\begin{document}

\title{Simulating quench dynamics on a digital quantum computer with data-driven error mitigation}

\author{Alejandro Sopena}
\author{Max Hunter Gordon}
\author{Germ\'{a}n Sierra}
\author{Esperanza L\'{o}pez}
\affiliation{Instituto de F\'{\i}sica Te\'{o}rica, UAM/CSIC, Universidad Aut\'{o}noma de Madrid, Madrid, Spain}
\begin{abstract}
Error mitigation is likely to be key in obtaining near term quantum advantage. In this work we present one of the first implementations of several Clifford data regression based methods which are used to mitigate the effect of noise in real quantum data. We explore the dynamics of the 1-D Ising model with transverse and longitudinal magnetic fields, highlighting signatures of confinement. We find in general Clifford data regression based techniques are advantageous in comparison with zero-noise extrapolation and obtain quantitative agreement with exact results for systems of $9$ qubits with circuit depths of up to $176$, involving hundreds of CNOT gates. This is the largest systems investigated so far in a study of this type. We also investigate the two-point correlation function and find the effect of noise on this more complicated observable can be mitigated using Clifford quantum circuit data highlighting the utility of these methods. 
\end{abstract} 

\maketitle

\section{Introduction}
The rapid progress in the field of quantum computing is encouraging, with current machines approaching the qubit qualities and system sizes expected to demonstrate some useful quantum advantage. However, noise within the computation still presents a large obstacle in obtaining useful results as current systems cannot implement full error correction. Therefore, it is expected that error mitigation techniques will be essential in demonstrating useful quantum advantage. These techniques aim to reduce the impact of noise rather than remove its effects completely. This relatively new field is experiencing a period of rapid progress with novel methods being developed in quick succession. Common approaches include quantum circuit compiling, machine learning~\cite{Murali,Cincio_2018, cincio2020machine} and variational algorithms~\cite{Peruzzo2014, cerezo2020variationalreview, sharma2019noise, O'Malley2016, cirstoiu2020variational}. Recent advances show phase estimation~\cite{o2020error} and so-called virtual state distillation~\cite{koczor2020exponential, huggins2020virtual, czarnik2021qubitefficient} can also be used for error mitigation and show great promise.

One of the most popular techniques is so called zero-noise extrapolation (ZNE)~\cite{Temme_2017}. Data from an observable of interest evaluated at several controlled noise levels is used to give an improved estimate of the noise free observable. Despite much success~\cite{otten2019recovering,  Dumitrescu_2018, Kandala_2019, cai2020multiexponential} this technique is limited by the assumption of low hardware noise, which may not be valid in the circuits of a size and depth necessary to demonstrate quantum advantage.

Recently, it has been shown that data sets produced by classically simulable quantum circuits such as near-Clifford circuits~\cite{czarnik2020error, strikis2020learning}, circuits based on fermionic linear optics or matchgate circuits~\cite{montanaro2021error} can be used to mitigate the effects of noise. In so called Clifford data regression (CDR) the exact and noisy data from near-Clifford circuits is used to learn a functional relation between the noisy and exact observables. This relation can then be applied to a noisy observable of interest which cannot be simulated classically. This technique can also be unified with ZNE~\cite{lowe2020unified}. In variable noise Clifford data regression (vnCDR) near-Clifford circuits are evaluated at several controlled noise levels. The exact and noisy values are then used to perform a guided extrapolation to the zero-noise limit. In general, these regression based methods appear advantageous over ZNE due to their simplicity and scalability. However, there are few examples of these methods being applied to real data from currently available quantum computers, where noise is significantly more difficult to mitigate.

One clear application of quantum technologies is the simulation of quantum systems. The classical resources necessary to simulate such systems in general scales exponentially with the system size. Spin - $\frac{1}{2}$ systems are particularly relevant as they map directly onto physical qubits, making spin chains an ideal testing ground for both current and future quantum computers~\cite{Smith_2019}. 
A common problem to consider in condensed matter simulations is non-equilibrium dynamics. These dynamics can be induced by a global quench, which is a sudden change to the system Hamiltonian.
Simulations obtaining quantitative accuracy have been reported in ion trap architectures~\cite{tan2019observation} and more recently in super-conducting architectures~\cite{vovrosh2021efficient}. 

In this work we provide a comparison of several error mitigation strategies applied to the problem of simulating a quantum quench in the one dimensional Ising model with transverse and longitudinal magnetic fields. To investigate how these methods perform in real quantum devices we explore the behavior of several observables of interest and simulate the system dynamics with various circuit depths. We measure the frequency of oscillations of the magnetisation for different initial states in a system of $9$ spins. Furthermore, we present one of the first measurements of the two-site correlation function in a study of these dynamics on a superconducting device. We are able to mitigate the effect of noise in data produced by deeper circuits and larger systems than previously explored in similar works. We find that Clifford regression based methods are able to obtain quantitative accuracy with the exact results and consistently outperform ZNE. 

First, we present an overview of the techniques used in data-driven error mitigation, with particular focus on CDR and vnCDR and their relation to recent advances. A slight variation of these methods is explored, namely "poor-man's" CDR (pmCDR), which performs well for short depth trotterised simulations of hamiltonian dynamics. We then review the theoretical expectations of the model and the methods used to implement the simulation in a super-conducting architecture. We show that using error mitigation we are able to obtain consistent quantitative accuracy involving circuits of depth $110$ with $160$ CNOT gates and $9$ qubits, while also obtaining some results with quantitative accuracy for depths up to $176$ with $256$ CNOT gates. Finally we conclude with a discussion of the results presented here and an outline of future directions.

\section{Data-driven error mitigation} 
Data-driven error mitigation uses classical post processing of quantum data to improve the zero-noise estimates of some observable of interest. In this work ZNE~\cite{Temme_2017}, CDR~\cite{czarnik2020error} and vnCDR~\cite{lowe2020unified} are used to obtain noise-free estimates of various observables. Furthermore, following the recent work showing the success of a simple mitigation strategy with an assumed noise model~\cite{vovrosh2021efficient}, we demonstrate the utility of a similar approach where the parameters of an assumed noise model are learned using near-Clifford circuits (pmCDR).

\subsection{ZNE}
Zero-noise extrapolation is one of the most popular error mitigation strategies. It uses quantum circuit data collected at various hardware noise levels to estimate the value of a noise free observable. Intuitively, by increasing the noise in a controlled manner and extrapolating to the zero-noise limit one can obtain a more accurate estimate of an observable of interest. Originally, this technique was presented within the context of stretching gate times to increase noise and using Richardson extrapolation to approach the zero-noise limit~\cite{Temme_2017}. More recently this has been extended to hardware agnostic approaches through unitary folding~\cite{giurgicatiron2020digital, larose2020mitiq} and identity insertion methods~\cite{He_2020}. Furthermore, additional extrapolation techniques have been proposed~\cite{PhysRevX.8.031027, otten2019recovering}. 

Despite widespread success ZNE performance guarantees are limited due to uncertainty in the extrapolation. Additionally, in real devices often the base-level noise is too strong to enable an accurate extrapolation, particularly in circuits with significant depth.

\subsection{CDR}
More recently, Clifford circuit quantum data has been used to mitigate the effect of noise~\cite{czarnik2020error, strikis2020learning}. Quantum circuits composed of mainly Clifford gates can be evaluated efficiently on a classical computer. In CDR near-Clifford circuits are used to construct a set of noisy and exact expectation values for some observable of interest. This dataset is used to train a simple linear ansatz mapping noisy to exact values. Following the presentation in~\cite{czarnik2020error} taking $\hat{\mu}_{0}$ to be the observable evaluated with hardware noise, CDR uses Clifford circuits to train the following anstaz:
\begin{equation}
f(\hat{\mu}_0) = a_1\hat{\mu}_0 + a_2
\label{eq:cdr_ansatz}
\end{equation}
The parameters $a_1$, $a_2$ are chosen using least-squares regression on the near-Clifford circuit dataset. For a training set of $m$ Clifford circuits with noisy expectation values $\{x_i\}$ and exact expectation values $\{y_i\}$ evaluated classically, one calculates 
\begin{equation}
    (a_1^{*},a_2^{*}) = \underset{(a_1,a_2)}{\text{argmin}} \sum_{i=1}^m \left[y_i - (a_1 x_i + a_2)\right]^2. 
\end{equation}
These learned parameters are then used to mitigate the effect of noise on an observable produced by a circuit which is not classically simulable. As noted in Ref.~\cite{czarnik2020error} the form of the anstaz can be motivated by considering the action of a global depolarizing channel. Letting $\rho$ be the density matrix for the noise-free state after some evolution.
Consider the action of a depolarizing noise channel $\mathcal{E}$ which acts on this state before a measurement of some observable of interest $X$. The action of the channel can be described as follows
\begin{align}
    \tr(\mathcal{E}(\rho) X) = (1-\epsilon)\tr(\rho X) + \frac{\epsilon\tr(X)}{d}
\label{eq:depolarising}
\end{align}
where $d$ is the dimension of the system and $\epsilon\in\qty(0,1)$ is a parameter characterizing the noise.  Identifying $\hat{\mu}_0 = \tr(\mathcal{E}(\rho) X)$ and
\begin{align}
    a_1 = 1/(1-\epsilon),\quad a_2 = -\frac{\epsilon}{d(1-\epsilon)}\tr(X)
\label{eq:clifford_parameters}
\end{align}
the noise-free expectation value $\tr(\rho X)$ can be calculated using Eq.~\eqref{eq:cdr_ansatz}. Therefore, in the case of a global depolarizing channel CDR should perfectly mitigate the noise, assuming the Clifford circuit training set accurately captures its effect. This ansatz also perfectly corrects certain types of measurement error~\cite{czarnik2020error}. 

For observables $X$ with $\tr(X) = 0$ the linear term in CDR appears to be redundant assuming the noise can be modelled with a global depolarising channel, shown to be an accurate description in some circumstances~\cite{vovrosh2021efficient}. However, we find including the constant term in the anstaz allows for more flexible fitting of the training data, leading to a better mitigation in general (\textit{e.g.} for the data shown in Fig.~\ref{fig:magnetisation} the absolute error is improved by a factor of $1.2$). An example of such a case can be seen in Appendix~\ref{sec:CDR_cnst}.

\subsection{Poor man's CDR}
\label{sec:pmcdr}
As previously mentioned, recently it has been shown that a global depolarising channel (Eq.~\ref{eq:depolarising}) appears to accurately describe the noise in a real device for small system sizes~\cite{vovrosh2021efficient, urbanek2021}. Indeed, this noise model provides the motivation for use of a linear anstaz in CDR. Here, we implement a simplified version of CDR where short depth near-Clifford circuits are used to fit the parameter $\epsilon$ characterising a global depolarising noise channel:
\begin{equation}
    \expval{X}_{noisy} = (1-\epsilon)\expval{X}_{exact} +\frac{\epsilon \tr(X)}{d}.
    \label{pmcdr_ansatz}
\end{equation}
Training sets are constructed from the quantum circuits of one and two Trotter steps. This data is used to determine $\epsilon_1$ and $\epsilon_{2}$ respectively. Due to the repetitive structure of the circuit in a trotterised evolution we assume the effect of the error on an observable can be modelled using Eq.~\eqref{pmcdr_ansatz}, with the parameters evolving as $(1-\epsilon)^{\alpha N_{T}}$, where $N_{T}$ is the number of Trotter steps and $\alpha$ is some constant (see also~\cite{vovrosh2021efficient}). Using $\epsilon_{1}$ and $\epsilon_{2}$, determined with the near-Clifford training data, we can fit $\alpha$ and use this assumed model to correct observables from circuits involving more Trotter steps. We find $\alpha$ is close to one for the magnetisation while for $\Delta_i^{ZZ}(t)$ it is higher (\textit{e.g.} the mean values of $\alpha$ are 1.01 and 1.30 for the magnetisation and $\Delta_i^{ZZ}(t)$ results shown in Figs.~\ref{fig:magnetisation} and.~\ref{fig:delta} respectively).

The advantage of this approach is that it is only necessary to produce two near-Clifford data sets. The parameters of the noise model are then learned and applied to observables from other circuits. This is more convenient, having a reduced experimental and computational overhead. We note a similar technique was recently presented using estimation circuits consisting of only CNOT gates and combining this idea with randomised compiling and zero-noise extrapolation~\cite{urbanek2021}.

\subsection{vnCDR}
Zero noise extrapolation and Clifford data regression can be conceptually unified into one mitigation strategy where Clifford circuit quantum data is used to inform the functional form of the extrapolation to the zero-noise limit~\cite{lowe2020unified}. Intuitively, so called variable noise Clifford data regression reduces the risk of blind extrapolation and is expected to outperform both ZNE and CDR in deep quantum circuits involving many qubits. vnCDR makes use of Clifford circuits evaluated at several noise levels to train a more general anstaz than that of CDR. Considering $m$ near-Clifford circuits and $n+1$ noise levels $c_j\in \mathcal{C}$, a noisy estimate of the observable expectation value is defined as $x_{i, j}$. For each of the $m$ circuits the corresponding exact observable $y_i$ is computed classically. The training set $\mathcal{T}$ is taken as $\mathcal{T} = \{(\vec{x}_i, y_i)\}$ where $\vec{x}_i = (x_{i,0},\dots,x_{i,n})$ is the vector of noisy expectation values produced by the $i^\text{th}$ circuit. This training data is used to learn a function that takes a set of noisy estimates at the $n+1$ different noise levels and outputs an estimate for the noise-free value. We use the linear ansatz
    \begin{equation}
       g(\vec{x}; \vec{a}, b) = \vec{a}\cdot\vec{x} + b \,,
    \end{equation}
where we have included a constant term $b$. Least-squares regression is used on the dataset $\mathcal{T}$ to pick optimal parameters $\vec{a}^*, b^*$, i.e.,
\begin{align}
    (\vec{a}^*, b^*) = \underset{\vec{a}, b}{\text{argmin}} \sum_{i=1}^m \left(y_i - (\vec{a}\cdot \vec{x}+ b)\right)^2\,.
\end{align}
Therefore, $g(\vec{x}; \vec{a}^*, b^*)$ is expected to output a good estimate for the noise-free expected value from a vector formed of the noisy expectation values at different controlled noise levels. This mitigation strategy is also expected to perfectly mitigate for a global depolarising noise channel. Despite promising results the performance of vnCDR has not been extensively explored on real quantum circuit data, motivating the analysis we present here.

Originally vnCDR was introduced with an ansatz excluding the constant term $b$, appearing more similar to  Richardson extrapolation~\cite{lowe2020unified}. For the observables we consider here including this parameter made for a more accurate mitigation (\textit{e.g.} for the data shown in Fig.~\ref{fig:magnetisation} the absolute error  is improved by a factor of $1.1$).

Overall, classically simulable near-Clifford circuits can be used to inform the experimenter about the noise present in the device. CDR makes use of extracting this data for every circuit to mitigate the results of an observable of interest from that particular circuit. Assuming a noise model, pmCDR makes use of two Clifford data sets and uses this data to complete a mitigation on circuits with a repetitive structure at different depths. Data collected at various artificial controlled noise rates also contains relevant information to perform a mitigation as shown in ZNE. vnCDR conceptually unifies ZNE and CDR by collecting near-Clifford circuit data at various noise levels. 

\section{Model}
Data-driven error mitigation is a promising approach to reduce the impact of noise in near term quantum computers. One of the areas where quantum algorithms are expected to show some advantage over classical methods is in simulating quantum many body systems. A system which displays interesting many body dynamics is the TFIM with an additional longitudinal field, providing a clear test bed for these mitigation methods.  
\subsection{Transverse-Longitudinal Ising model}
The Hamiltonian of the quantum one-dimensional Ising model of length $L$ with transverse and longitudinal fields is given by
\begin{equation}\label{IisingHamiltonian}
H=-J\left[\sum_{i}^{L} \hat{\sigma}_{i}^{Z} \hat{\sigma}_{i+1}^{Z}+h_{X} \sum_{i=1}^{L} \hat{\sigma}_{i}^{X}+h_{Z} \sum_{i=1}^{L} \hat{\sigma}_{i}^{Z}\right]
\end{equation}
where $J$ is an exchange coupling constant, which sets the microscopic energy scale and $h_X$ and $h_Z$ are the transverse and longitudinal relative field strengths, respectively. This model is integrable for $h_Z=0$ while for $h_Z\neq0$ it is only integrable in the continuum when $h_X=1$~\cite{zamolodchikov_integrals_1989}.

Setting $h_Z=0$, in the continuum limit, the diagonalisation of the Hamiltonian results in the description of a fermion with mass $m=2J\abs{1-h_X}$ and velocity $v=2J\sqrt{h_X}a$ where $a$ is the chain spacing and $ka\ll1$~\cite{sachdev_quantum_2011}. At $h_X=1$, the system has a critical point and the low-energy behaviour of the system is described by a conformal field theory with central charge $c=1/2$~\cite{mussardo_statistical_2010}. For $h_X<1$, the system is in the ferromagnetic phase (with $J>0$). This system can be approximated by considering the low-energy elementary particle excitations which are given by domain walls between the two ground states of $H$ with $h_X=0$~\cite{sachdev_quantum_2011},
\begin{equation}
\ket{i}=\ket{\uparrow\cdots\uparrow_{i-1}\uparrow_{i}\downarrow_{i+1}\downarrow_{i+2}\downarrow_{i+3} \cdots\downarrow}.
\end{equation}
These states are identified as fermions.

A longitudinal field $h_Z\neq0$ induces a confining potential between pairs of domain walls,
\begin{equation}
\ket{i,n}=\ket{\uparrow\ldots\uparrow_{i-1}\downarrow_{i} \ldots \downarrow_{i+n-1} \uparrow_{i+n} \ldots  \uparrow},
\label{eq:meson}
\end{equation}
which increases linearly with the length of the domain, $n$. This leads to excitations formed from pairs of domain walls Eq.~\eqref{eq:meson}, which are referred to as mesons~\cite{McCoy:1994zi}.

In order to show the temporal evolution of the position of fermions and mesons, we measure the probability distribution of kinks,
\begin{equation}
    \Delta_i^{ZZ}(t)=\frac{1}{2}(1-\expval{\hat{\sigma}_i^Z\hat{\sigma}_{i+1}^Z}),
\end{equation} 
from an initial state of $2$ kinks (see Appendix~\ref{sec:masses}). This observable takes the value $0$ when there are no kinks and is $1$ when the $i$-th and $(i+1)$-th spins form a kink.

Confinement suppresses the light cone spreading of correlations~\cite{kormos_real-time_2017}. This effect can be seen by measuring the two point correlation function,
\begin{equation}
    \sigma_{i, j}^{Z Z}(t)=\expval{\hat{\sigma}_i^Z\hat{\sigma}_j^Z}-\expval{\hat{\sigma}^Z_{i}}\expval{\hat{\sigma}_j^Z}.
\end{equation}

In the presence of a longitudinal field and $h_X<1$ local observables after quenches exhibit oscillations whose dominant frequencies are the energy gaps between bound states~\cite{kormos_real-time_2017} (see Appendix~\ref{sec:masses}). These energy gaps can be interpreted as meson masses. A suitable observable for measuring the meson masses is the magnetisation, $\sigma_i^Z(t)=\expval{\hat{\sigma}_i^Z}$. In order to avoid edge effects we measure the magnetisation at the centre of the chain for initial states without domain walls and at the outer edge of the domain wall for initial states of two domain walls~\cite{tan2019observation}. 

In this work we explore the signatures of confinement by measuring the probability distribution of kinks $\Delta_i^{ZZ}(t)$, the evolution of the two point correlation function $\sigma_{i, j}^{Z Z}(t)$ and the meson masses determined by extracting the dominant frequency of the oscillation of the magnetisation $\sigma_{i}^Z$. 

\subsection{Quantum simulation}
We simulate the induced Hamiltonian dynamics using a first order trotterised evolution of the initial state. We start by discretising the evolution operator in $n$ blocks such that
\begin{equation}
    U(t)=e^{-iHt}=\qty(e^{-iH\Delta t})^n=U^n\qty(\Delta t)
\end{equation}
with $\Delta t=t/n$. Each evolution step operator is approximated using the first order Trotter expansion:
\begin{equation}
    U(\Delta t)=e^{-iH\Delta t}=\prod_{k} e^{-i h_{k} t / n}+\mathcal{O}((\Delta t)^{2})
\end{equation}
where $h_k=-J\left[ \hat{\sigma}_{k}^{Z} \hat{\sigma}_{k+1}^{Z}+h_{X}  \hat{\sigma}_{k}^{X}+h_{Z}  \hat{\sigma}_{k}^{Z}\right]$. To implement $e^{-i h_{k} t / n}$ on IBM devices, we decompose the quantum circuit to execute one Trotter step into the native IBM gate set  $\{R_{X}(\pi/2), R_{Z}(\theta), X, \text{CNOT}\}$ (see Appendix~\ref{sec:circuit}). This decomposition leads to a depth of $11$ per Trotter step with $2(Q-1)$ CNOT gates for a system size $Q>2$, where $Q$ is the number of qubits. For a fixed time step $\delta t$ one can evaluate the dynamics up to time $t$ by repeated action of this circuit $N_{T}$ times, where $N_{T}=t/\delta t$.

\begin{figure*}[htb!]
    \centering
    \includegraphics[width = 0.99\textwidth]{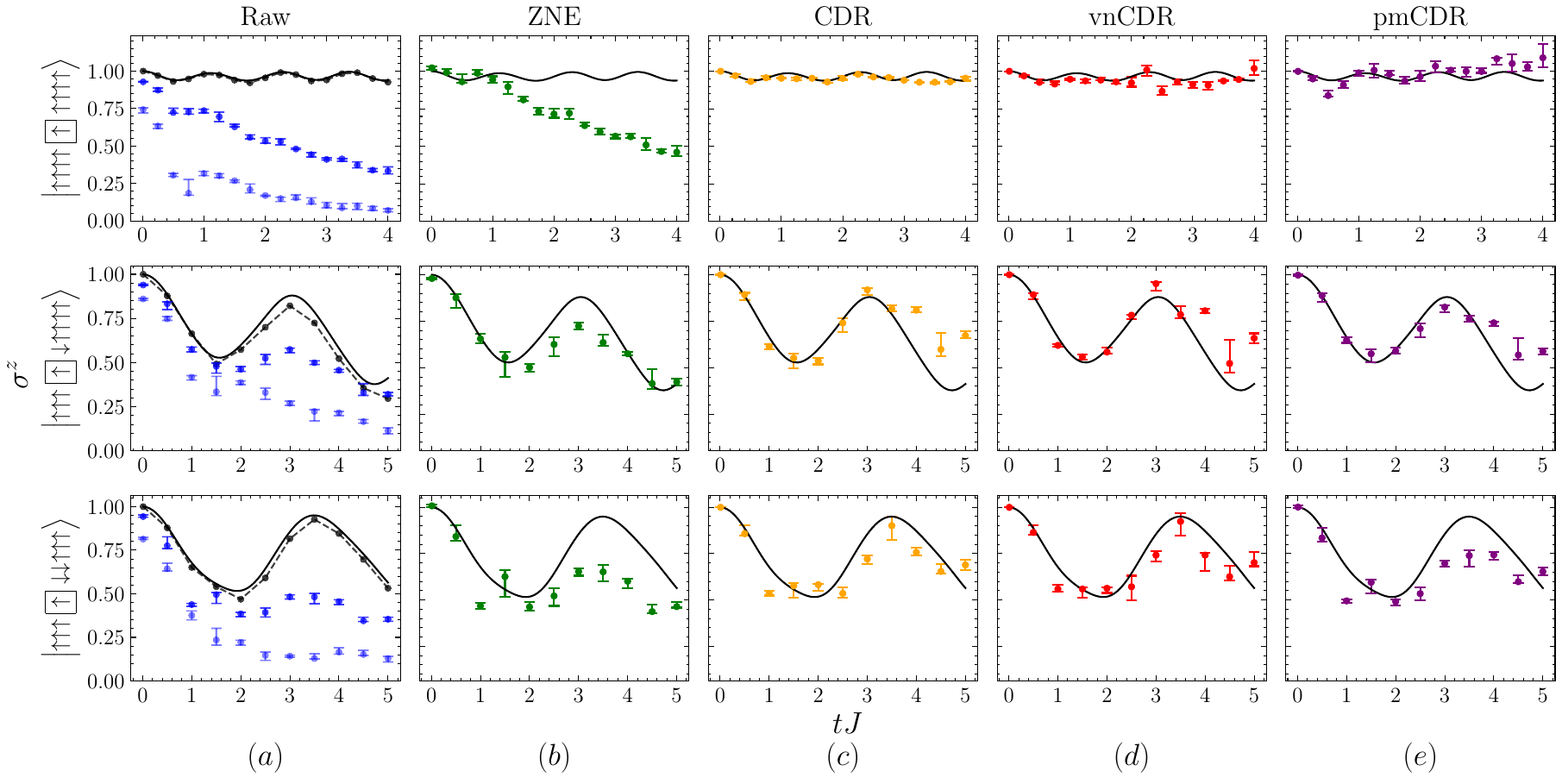}
    \caption{Temporal evolution of $Z$-axis local magnetisation with $h_X=0.5$ and $h_Z=0.9$ for three initial states (in each row) with the observables mitigated using various techniques (columns). In all panels the exact diagonalised dynamics is shown  as a black-solid line. Raw observables are shown in the left most column ($a$) calculated at two noise levels $\mathcal{C} = \{1, 3\}$ (blue and light blue points respectively) using the IBMQ Paris quantum computer. Black points and dashed lines show the trotterised dynamics. Error bars show the distribution of observables calculated over $6$ repeats of the circuits of interest, and central points show the median. The raw observables at the first noise level were calculated to have a mean absolute error of $0.279$. ZNE reduced this error to $0.166$, CDR to $0.094$, vnCDR to $0.092$ and pmCDR to $0.096$.}
    \label{fig:magnetisation}
\end{figure*}

\section{Simulated Spin chain confinement}
In this section we display the results obtained after applying the mitigation methods described above on the trotterised evolution of a system of $Q=9$ qubits. We investigate three observables of interest: the magnetisation, $\sigma^{Z}_i$, to determine the masses of the mesons, $\Delta^{ZZ}_i$, to visually demonstrate confinement and two-point correlation function, $\sigma^{ZZ}_{ij}$, to explore how the mitigation techniques perform on a more complex, non-local observable. Every circuit used, both in training set construction for mitigation and in collecting raw data, was evaluated with $8192$ shots. We use the absolute error to quantitatively explore the performance of the mitigation strategies implemented:
\begin{equation}
    error = \left|\frac{\langle X \rangle_{mitigated} - \langle X \rangle_{trotterised}}{mean(\langle X \rangle_{trotterised})}\right|,
\end{equation}
where the mean is taken over the time evolution. 

\subsection{Local magnetisation evolution}

To  determine the first meson masses we measure the oscillations of the local magnetisation for three initial states. We extract the dominant frequencies using a single-frequency sinusoidal fit as in Ref.~\cite{vovrosh2021efficient},
\begin{equation}
    \sigma^Z_j=a_1e^{-a_2t}\cos(a_3t)+a_4t+a_5.
\label{eq:fit}
\end{equation}

We explore the evolution starting with the system initialised as: all spins up, the central qubit in the down state and all other qubits up and two central qubits in the down state with all others spin up. The system is evolved using a trotterised evolution of the Hamiltonian for a fixed time step. The circuit depth therefore grows linearly with the number of Trotter steps. When the system is initialised to all spins up, the Trotter step was chosen to be $0.25J$ leading to a final circuit depth of $176$ with $256$ CNOT gates. In both other initialisations the Trotter step was $0.5J$ leading to a final circuit depth of $110$ involving $160$ CNOT gates. A different Trotter step is needed in these two cases to reproduce the smaller amplitude and higher frequency oscillations observed with the initial state being all spins up. 

The temporal evolution of the magnetisation is shown in Fig.~\ref{fig:magnetisation}. The raw values for the magnetisation clearly decay towards the maximally mixed state with circuit depth. With a higher noise level this can be seen to occur more quickly, as expected. We mitigate the raw results using ZNE, CDR, vnCDR and pmCDR.

\begin{figure}[htb!]
    \centering
    \includegraphics[width = 0.99\linewidth]{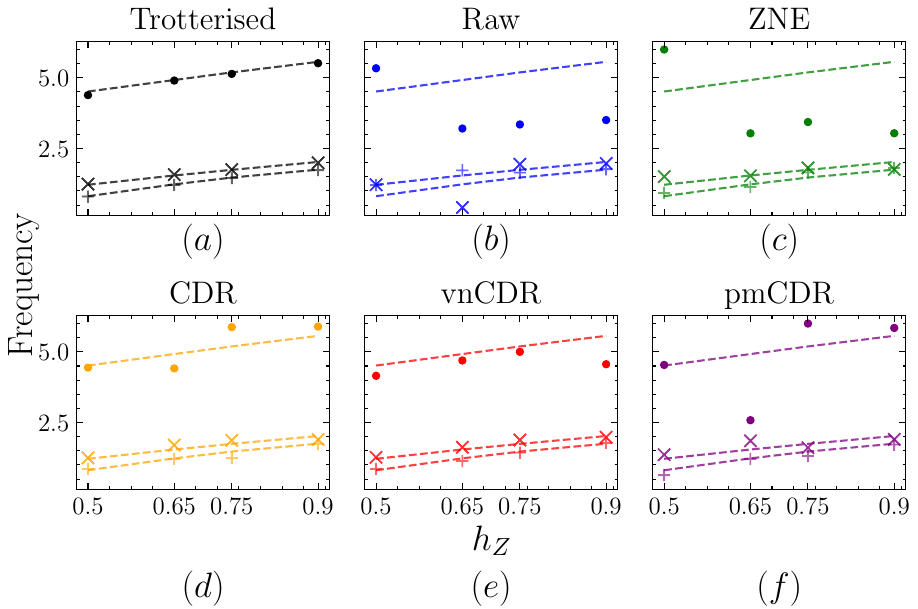}
    \caption{Frequencies obtained at $h_X=0.5$ and various $h_{Z}$ values calculated from the exact diagonalised (dashed lines), trotterised (a) and the median raw (b) and median mitigated results (c)-(f). Frequencies obtained for initial states $\ket{\uparrow\uparrow\uparrow\uparrow\uparrow\uparrow\uparrow\uparrow\uparrow}$ (dots),  $\ket{\uparrow\uparrow\uparrow\uparrow\downarrow\uparrow\uparrow\uparrow\uparrow}$ (diagonal cross) and  $\ket{\uparrow\uparrow\uparrow\downarrow\downarrow\uparrow\uparrow\uparrow\uparrow}$ (vertical cross) are plotted.}
    \label{fig:freq_0.5}
\end{figure}

From the time evolution of the magnetisation for different values of $h_z$ (see Appendix~\ref{sec:data}) we obtain the dominant frequencies shown in Fig.~\ref{fig:freq_0.5}. In order to calculate the frequencies it is not necessary to fit the entire evolution of the magnetisation. We found more accurate values are obtained by fitting times up to around $tJ=3$.

\begin{figure*}[htb!]
    \centering
    \includegraphics[width = 0.99\textwidth]{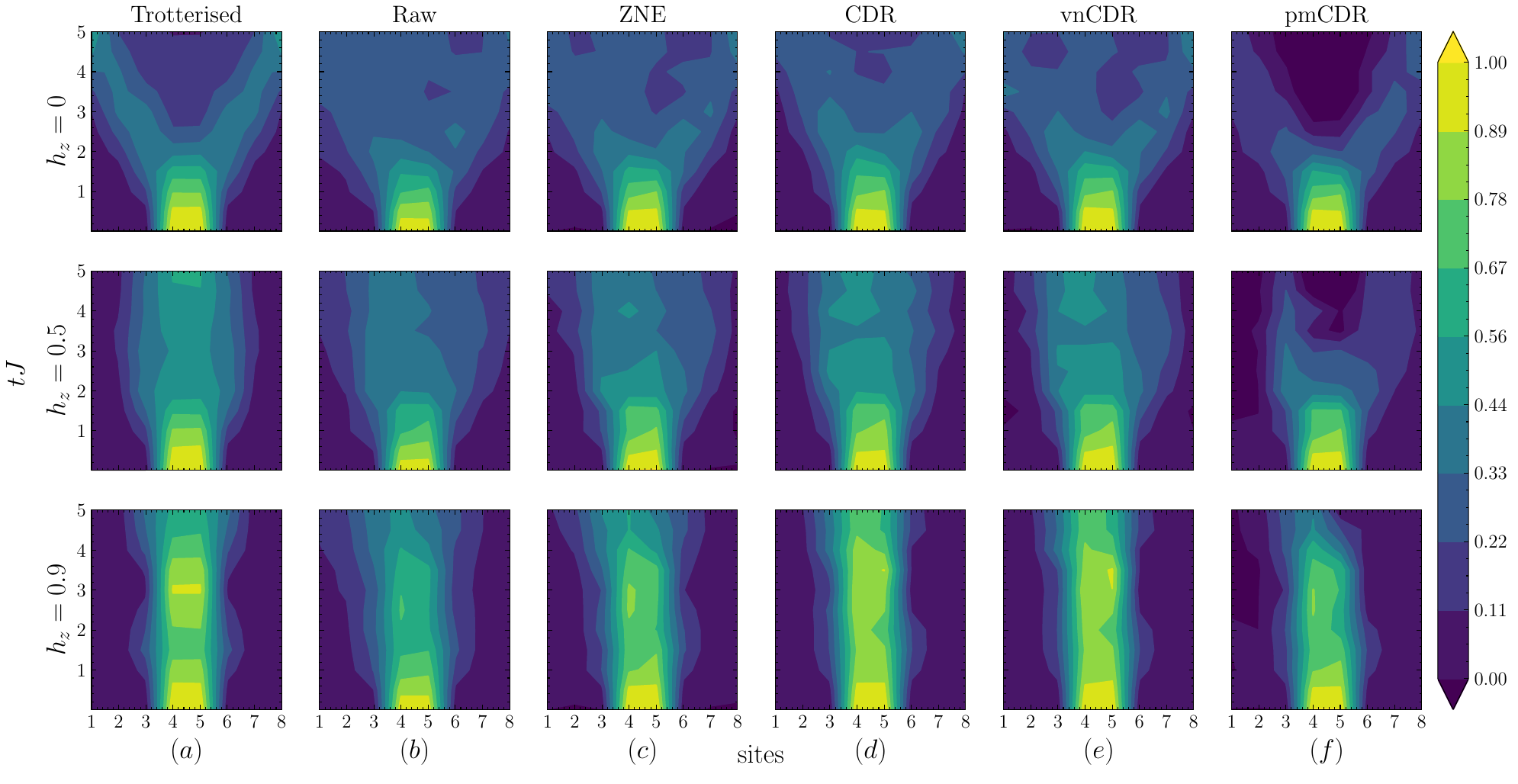}
    \caption{The observable $\Delta_{i}^{ZZ}$ projected into the 2-kink subspace, measured at various sites and Trotter steps when $h_{X} = 0.5$, $h_{Z} = 0$ (upper row), $h_{X} = 0.5$, $h_{Z} = 0.5$ (middle row) and $h_{X} = 0.5$, $h_{Z} = 0.9$ (bottom row) . The initial state of the system is $\ket{\uparrow\uparrow\uparrow\uparrow\downarrow\uparrow\uparrow\uparrow\uparrow}$. We mitigate the raw observables using ZNE ($c$), CDR ($d$), vnCDR ($e$) and pmCDR ($f$). The raw observables have a mean absolute error of $0.300$. ZNE reduces this error to $0.197$, CDR to $0.154$, vnCDR to $0.147$ and pmCDR increases the absolute error to $0.410$.}
    \label{fig:delta}
\end{figure*}

All mitigation methods improve upon the raw data. In particular CDR, vnCDR and pmCDR mitigate the effect of noise effectively in many cases, even for deep circuits. In some cases, like those shown in Fig.~\ref{fig:freq_0.5}, pmCDR performs as well as the other Clifford based methods. This suggests due to the repetitive structure of the quantum circuit the effect of noise on this observable can be characterised more easily, using only the first two Trotter steps. However, we find this is not as reliable as using a training set to learn the noise at each Trotter step, as done in CDR and vnCDR. This indicates the noise parameters can vary considerably beyond some circuit depth and also in short time frames, between runs. Still, it is quite remarkable that a simple global noise model describes the noise so accurately in some runs (see Appendix~\ref{sec:data} for more examples). 

For more complicated observables treating the noise as purely depolarising breaks down more quickly and pmCDR begins to perform worse. It should be noted that we do not implement measurement error mitigation. While we do not expect this to impact the performance of CDR or vnCDR it should lead to worse implementations of both pmCDR and ZNE. We also do not enforce any physical constraints on our mitigated observables, in order to asses the raw potential of the method. Therefore, occasionally pmCDR gives unphysical values, increasing the absolute error of the mitigation significantly. 

Overall, CDR and vnCDR show the most reliable mitigation of the magnetisation. They consistently offer a quantitatively accurate mitigation for times up to $tJ\sim5$ and occasionally up to even longer times. Therefore, it can be concluded the computational overhead necessary for these methods is useful in mitigating the effect of noise. vnCDR does not offer any significant visual advantage over CDR although it does lead to the most accurate calculations of the frequencies and often has a smaller absolute error on average. ZNE and pmCDR perform consistently well for shorter depth circuits. 

\subsection{Kink evolution}
Starting with the initial state $\ket{\uparrow\uparrow\uparrow\downarrow\downarrow\uparrow\uparrow\uparrow\uparrow}$, projecting into the two kink subspace and measuring the observable $\Delta_{i}^{ZZ}$ shows the evolution of the kinks with time. Confinement can be directly observed by comparing the evolution when $h_{Z} = 0$ and $h_{Z} \neq 0$. With the presence of some transverse field we observe an attraction between the kinks due to the confining potential. These dynamics are more difficult to mitigate in comparison with the local magnetisation as to observe this effect the entire state of the system is probed. This gives a good indication of average performance of each method. In Fig.~\ref{fig:delta} we show the evolution of $\Delta_{i}^{ZZ}$ with the various mitigation methods for $h_{X} = 0.5$ and $h_{Z} = 0,\text{ }0.5,\text{ }0.9$. We note that when $h_{Z}=0$ there are $9$ less non-Clifford gates per Trotter step. 

The oscillations that appear to be washed out in the raw data are clearly recovered by CDR and vnCDR. ZNE does partially recover the oscillations but not to the same accuracy. It is important to note that pmCDR begins to fail at around $tJ = 3$ leading to an increased absolute error in comparison with the raw results.

In the pmCDR implementation since $\tr\qty(\Delta^{ZZ})\neq0$, the ansatz relating the mitigated and the noisy observables Eq.~\eqref{pmcdr_ansatz} has a linear term and a constant term dependent on the parameter $\epsilon$. Therefore, to obtain an error of the same magnitude as that of the magnetisation, the difference between the true $\epsilon$ parameter and the one obtained from the fit must be smaller for $\Delta_{i}^{ZZ}$ than for the magnetisation. If during operation the true noise model changes slightly this has a large impact on the results. In addition, because $\Delta_{i}^{ZZ}$ is a two-qubit observable, the ansatz is perhaps too simple to fully characterise the noise. The impact of measurement error is also detrimental. Furthermore, we note the results from pmCDR could be improved by enforcing physical constraints on the mitigated values.

Without mitigation the dynamics of the observables is not significantly changed by the introduction of a transverse field. The separation velocity does appear to be reduced but no oscillatory dynamics is observed. However, with CDR and vnCDR these oscillations are recovered and there is a striking visual contrast between the dynamics with and without the presence of a longitudinal field. We deduce from these results that CDR and vnCDR appear to be the more powerful mitigation strategies. 

\subsection{Correlation evolution}
We also investigate the correlation with the central qubit as the system evolves. This observable is non-local and is formed by combining three observables $\expval{\hat{\sigma}_i^Z\hat{\sigma}_5^Z}$, $\expval{\hat{\sigma}^Z_{i}}$ and $\expval{\hat{\sigma}^Z_{5}}$, which we mitigate separately before combining. In general, the correlation decreases with $h_Z$ while $\expval{\hat{\sigma}_i^Z\hat{\sigma}_5^Z}$ and $\expval{\hat{\sigma}^Z_{5}}$ increase. As the longitudinal field increases, it becomes more complicated to mitigate the correlation since the difference between the values of the two correlation terms needs to be smaller. Therefore, the mitigation needs to perform very effectively on each term and the correlation proves challenging to mitigate for general values of $h_{Z}$ and $h_{X}$. Thus, we focus on the case with $h_Z < h_X$. This together with the edge effects due to the finite size of the system results in values for the correlation with $h_Z\neq0$ similar to values obtained with $h_Z=0$. 

\begin{figure}[htb!]
    \centering
    \includegraphics[width = 0.99\linewidth]{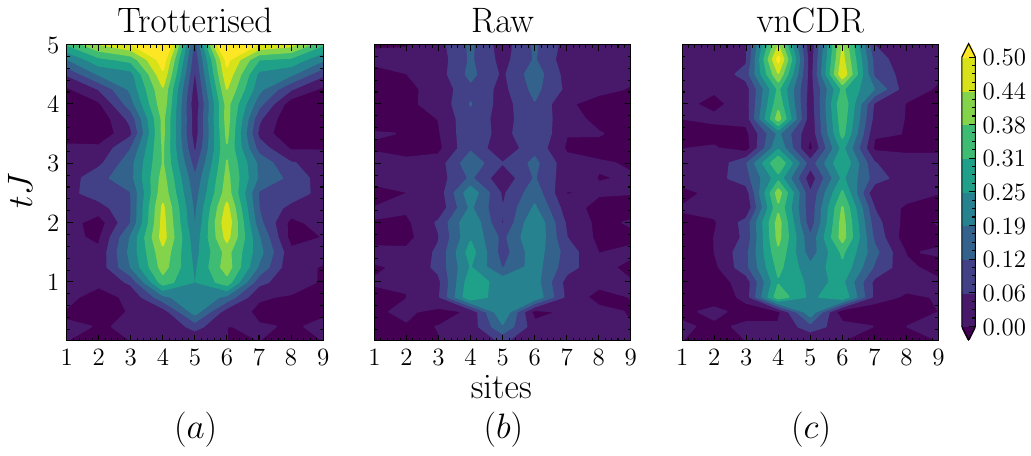}
    \caption{Correlation of qubits at sites along the x axis with the central qubit for the TFIM with $h_Z=0$ and the system initialised in the $\ket{\uparrow\uparrow\uparrow\uparrow\uparrow\uparrow\uparrow\uparrow\uparrow}$ state. The mean absolute error relative to the trotterised dynamics for the raw observables is $0.716$ , normalised by the mean value for the correlation across all times. vnCDR reduces this error to $0.457$. In this case vnCDR gave the best mitigated values for the correlation closely followed by CDR.}
    \label{fig:correlation_0}
\end{figure}

However, CDR and vnCDR do provide an advantageous mitigation with impressive visual results in some cases as shown in Fig.~\ref{fig:correlation_0} where we exhibit the correlation at $h_X=0.9$ and $h_Z=0$ with a Trotter step of $0.25$ which gives a final depth of $220$ at $tJ=5$, involving $320$ CNOT gates. The correlation with $h_Z=0.2$ is shown in Appendix~\ref{sec:data}. The dynamics which are almost entirely lost at late times are recovered to qualitative accuracy by vnCDR. In this case vnCDR mitigated results gave the lowest normalised absolute error. CDR also performed well giving very similar visual results. Showing that the dynamics of a complex observable can be qualitatively recovered for such deep circuits is a testament to the power of CDR and vnCDR.

Overall, we find that CDR and vnCDR lead to the best mitigated results for the observables explored in this work in almost all cases. The advantage is particularly clear for more complex observables. Interestingly, although vnCDR does generally have the smallest absolute error the advantage over CDR is slight. This could be attributed to how noise is being increased in the circuits of interest and when constructing the training set. More fine grained methods such as random identity insertions~\cite{He_2020} may be necessary to obtain a clear contrast between CDR and vnCDR. Alternatively, the lack of advantage in using multiple noise levels and near-Clifford training data could be due to the number of shots used to evaluate each circuit. More computational overhead might be necessary to obtain some improvement in the vnCDR results in comparison with CDR.

\section{Implementation details}
\subsection{Scaling the noise}
We perform the noise amplification in our quantum circuits using the so called fixed identity insertion method~\cite{He_2020}. We insert pairs of CNOT gates, which evaluate to the identity, after each CNOT implementation in the original circuit. Assuming the vast majority of error is introduced by these entangling gates, this method amplifies the noise by the factor of CNOT gates introduced. In our experiments we found it optimal to use noise levels $\mathcal{C} = \{1,3\}$ when using ZNE and vnCDR. Furthermore, a linear fit was used to extrapolate to the zero noise limit. We note that it would be interesting to implement a more fine grained noise amplification technique to explore if the results obtained by ZNE and vnCDR could improve. Additionally, more complex functions could be used to execute the extrapolation. 

\begin{figure}[htb!]
    \centering
    \includegraphics[width = 0.49\textwidth]{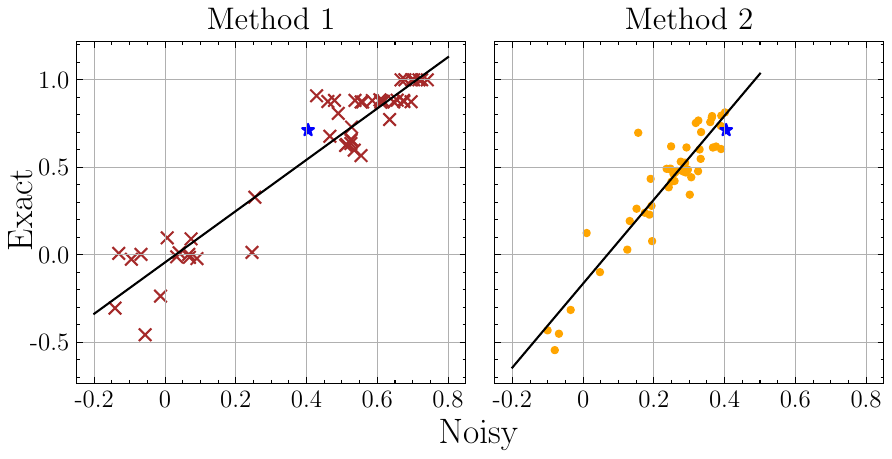}
    \caption{Distribution of exact and noisy magnetisation produced by the near-Clifford training circuits constructed using two different methods for time $t = 4$ ($8$ Trotter steps) with $h_X=0.5$, $h_Z=0.9$ and the initial state $\ket{\uparrow\uparrow\uparrow\uparrow\downarrow\downarrow\uparrow\uparrow\uparrow}$. Method $1$ refers to using probabilistic replacements throughout the entire circuit. Method $2$ refers to when every non-Clifford gate is fixed to appear after some circuit depth. The blue star shows the noisy and exact result for the observable from the circuit of interest.}
    \label{fig:training_set_comparison1}
\end{figure}

\begin{figure*}[htb!]
    \centering
    \includegraphics[width = 0.95\textwidth]{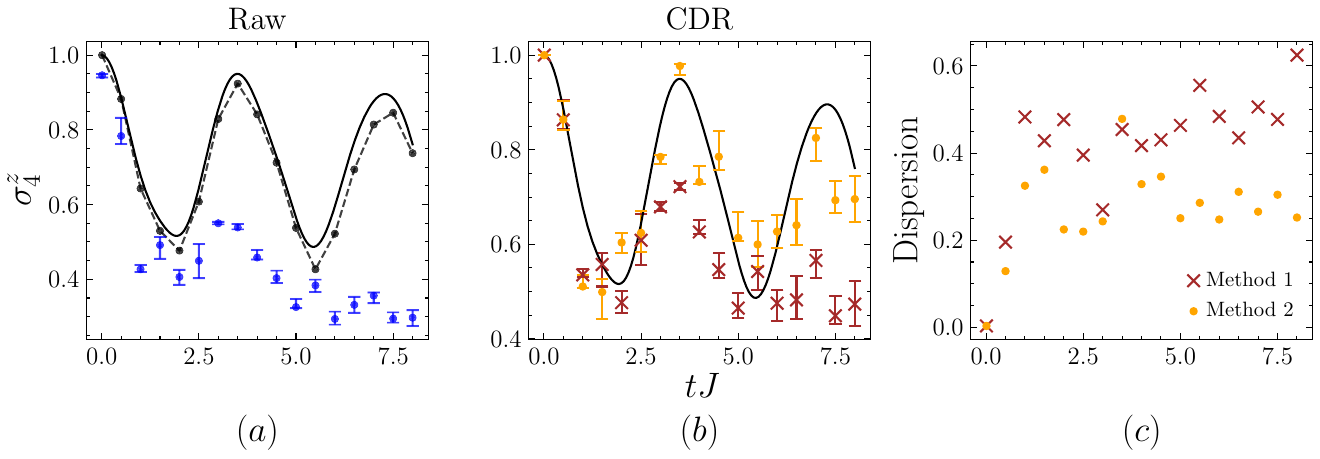}
    \caption{Evolution of the $\ket{\uparrow\uparrow\uparrow\downarrow\downarrow\uparrow\uparrow\uparrow\uparrow}$ state with $h_X=0.5$ and $h_{Z} = 0.9$. Exact diagonalised (black-solid), Trotterisied (black-dashed) and raw results shown in $(a)$. Data mitigated using CDR with two training set construction methods is shown in $(b)$, where brown crosses show the results from mitigating with a training set constructed with method $1$ and orange points show results using method $2$. The final circuit depth is 176 with 216 CNOT gates. Error bars show distribution of six repeats of the circuit of interest. The dispersion of each training set constructed by both methods at various circuit depths is shown in $(c)$.}
    \label{fig:training_set_comparison2}
\end{figure*}

\subsection{Near-Clifford circuit training set}
Constructing the set of circuits that make up the training set is a key feature of both CDR and vnCDR. Intuitively, one desires a set of circuits close, in some sense, to the circuit of interest while also being diverse enough to accurately train the ansatz. In order to construct such a set of circuits we follow the protocol presented in Refs.~\cite{czarnik2020error, lowe2020unified}. In this work we restrict substitutions to a portion of the circuit beyond some depth. 

First the circuit of interest is decomposed into the native gate set of the IBM quantum computers $\{R_{X}(\pi/2), R_{Z}(\theta), X, \text{CNOT}\}$. These gates are Clifford with the exception of $R_{Z}(\theta)$ which is only Clifford when $\theta = n \pi/2$, where $n = 0,1,2,3$ and correspond to the phase gate $S^{n}$. Therefore, we replace some of the $R_{Z}(\theta)$ gates by the phase gate to some power $n$. 

Which gates (labelled $i$) to replace are chosen probabilistically according to distribution,
\begin{equation}
     p(\theta_{i}) \propto \sum_{n=0}^3\exp(-||R_{Z}(\theta_{i})-S^{n}||^{2}/\sigma^{2}),
\end{equation}
where $||.||$ represents the Frobenius norm and sigma is a constant parameter taken here as $\sigma = 0.5$. Additionally, which Clifford gate to replace a chosen $R_{Z}(\theta)$ rotation with is also chosen probabilistically,
\begin{equation}
     p'(n) \propto \exp(-||R_{Z}(\theta_{i})-S^{n}||^{2}/\sigma^{2}),
\end{equation}
also with $\sigma = 0.5$. We find this choice of $\sigma$ allows for construction of training sets which are diverse yet biased to the circuit of interest.

In both CDR and vnCDR implementations $50$ near Clifford circuits were constructed in this manner for each circuit of interest. Half the non-Clifford gates in each circuit were substituted, capped at $50$ non-Clifford gates.

Two approaches were compared: replacing gates throughout the entire circuit (method 1) and restricting the replacements to appear beyond a certain depth (method 2). We found that method 2 produces more similar observables to the circuit of interest, while still being sufficiently diverse. An example of a two CDR training sets constructed with both methods is shown in Fig.~\ref{fig:training_set_comparison1}. This example reflects the general trend observed, with training circuits being more similar to the circuit of interest when restricting Clifford substitutions to a fixed portion of the circuit. This kind of training set leads to a better mitigation for deeper circuits (see Fig.~\ref{fig:training_set_comparison2}).

This can be motivated by visualising a Clifford replacement as a unitary transformation on the original circuit. To minimise the action of this unitary one can imagine naively maximising the section of the circuit left unchanged, so forcing the Clifford substitution to appear as late as possible. We replace all non-Clifford gates in the second half of the circuit up to $50$ non-Clifford gates. Beyond $50$ non-Clifford gates we restrict all Clifford substitutions to appear at the greatest possible circuit depth. Fig.~\ref{fig:training_set_comparison2}$(c)$ shows the dispersion of each training set constructed by both methods at various circuit depths. We use a measure of dispersion to indicate the closeness of the training circuits to the circuit of interest, defined as:
\begin{equation}
    \frac{1}{m} \sum_{i}^{m} \sqrt{(x_{i}-\langle X \rangle_{noisy})^{2} + (y_{i} - \langle X\rangle_{exact})^{2}},
\end{equation}
where $m$ is the number of training circuits and $x_{i}$, $y_{i}$ are the noisy, exact expectation values for the observable of interest for each of the training circuits and $\langle X \rangle_{noisy}$, $\langle X \rangle_{exact}$ are the noisy, exact expectation values for the circuit of interest. 

In Fig.~\ref{fig:training_set_comparison2}$(b)$ an example of CDR is shown successfully mitigating noise in deep circuits, with this figure showing the dynamics of the magnetisation being recovered up to the final circuit depth of $176$. Oscillations in the magnetisation are recovered after they all but vanish from the raw data. The dispersion increases less quickly with circuit depth for method 2 than for method 1, shown in Fig.~\ref{fig:training_set_comparison2}$(c)$, suggesting method 2 makes for more reliable training sets. This is reflected in the more accurate mitigation results obtained. 

In the case of pmCDR the training sets for the first two Trotter steps were used to train the model as outlined in Section \ref{sec:pmcdr}.

Once the circuits in the training set are executed (at two noise levels for each circuit of interest) this data is used to train the CDR and vnCDR ansatzes. We found for the majority of the observables investigated here the mitigation improved by repeatedly training the given anstaz on a randomly selected subset of the total training data. We used $200$ subsets with data from $5$ circuits each, taking the final mitigation as the median mitigated observables produced from each subset. We leave systematic investigation of this bootstrapped training method for a later work. All the observables of interest here can be calculated from the counts measured in the $Z$ basis. Therefore, data from the same training set from each circuit could be used to mitigate the noise on all the observables of interest.

\section{Conclusion}
In this work we have simulated the dynamics of a quantum quench on the TFIM using a trotterised evolution on a quantum computer. We applied several data-driven error mitigation techniques, as well as presenting a simplified implementation of CDR, so-called pmCDR inspired by Ref.~\cite{vovrosh2021efficient}. Using these techniques we have shown it is possible to calculate the first meson masses with quantitative accuracy for systems of $9$ qubits, the largest system explored in a study of this type. Clifford based mitigation methods show the best performance overall. We have demonstrated quantitative accuracy can be obtained using CDR and vnCDR from observables produced by circuits with depths of up to $176$ involving hundreds of CNOT gates. Furthermore, we have shown CDR and vnCDR enable the recovery of dynamics which appear completely washed out due to noise, highlighted in our measurements of the observable $\Delta_{i}^{ZZ}$ and the two-site correlation. pmCDR does work well consistently for shorter depth circuits, but begins to struggle as depth increases. A similar trend is observed for ZNE. Combining pmCDR with other mitigation strategies such as  measurement error mitigation~\cite{funcke2020measurement}, random compilation and ZNE for the estimation of the noise parameters could improve its performance~\cite{urbanek2021}. In general CDR and vnCDR are advantageous due to the more general ansatzes fitted with training data which reflect the noise acting on the circuit of interest more accurately. 

We have shown that making Clifford substitutions in a fixed region of the circuit of interest, beyond some depth, makes for a more accurate mitigation. The best training set construction method to use in general is still an open question. Clifford circuits are clearly useful mitigation strategies, but their performance could be enhanced with the development of well studied methods to construct a faithful training set. Furthermore, the exploration of more complex ansatzes is sure to provide promise in mitigating noise, as well as using training data suited to specific problems~\cite{montanaro2021error}. Finally, the combination of these methods with more recent error mitigation advances such as virtual distillation appears to be a promising research direction~\cite{czarnik2021qubitefficient}. It would also be interesting to explore recent variational algorithms~\cite{gibbs2021longtime} in conjunction with Clifford circuit based error mitigation to obtain some computationally non-trivial results. 

Overall, improvement in quality of available mitigation techniques and quantum hardware becoming more widely accessible opens the possibility of near term useful quantum advantage. Near-Clifford circuit based mitigation methods are demonstrating their potential to become the staple error mitigation technique. 

\section*{Acknowledgments}
We thank Piotr Czarnik for useful discussions. We also thank the IBM Quantum team for making devices available via the IBM Quantum Experience. The access to the IBM Quantum Experience has been provided by the CSIC IBM Q Hub. A.S.G is supported by the Spanish Ministry of Science and Innovation under grant number SEV-2016-0597-19-4. M.H.G is supported by “la Caixa” Foundation (ID 100010434),
Grant No. LCF/BQ/DI19/11730056. This work has also been financed by the Spanish grants PGC2018-095862-B-C21, QUITEMAD+ S2013/ICE-2801, SEV-2016-0597 of the ”Centro de Excelencia Severo Ochoa” Programme and the CSIC Research Platform on Quantum Technologies PTI-001.

\bibliographystyle{unsrtnat}
\bibliography{main.bib}
\appendix
\section{CDR training set example\label{sec:CDR_cnst}}
\begin{figure}[htb!]
    \centering
    \includegraphics[width = 0.3\textwidth]{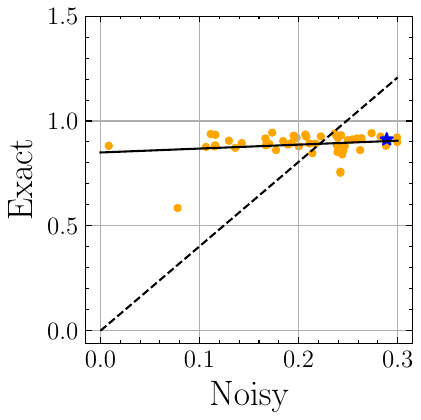}
    \caption{Distribution of exact and noisy magnetisation produced by the near-Clifford training circuits constructed for time $t = 3.75$ ($15$ Trotter steps) with $h_x=0.5$, $h_z=0.5$ and the initial state $\ket{\uparrow\uparrow\uparrow\uparrow\uparrow\uparrow\uparrow\uparrow\uparrow}$. The blue star shows the noisy and exact result for the observable from the circuit of interest. The continuous and dashed lines correspond to the ansatz Eq.~\eqref{eq:cdr_ansatz} with $a_2\neq0$ and $a_2=0$, respectively. We see that the constant term $a_2$ provides a clear advantage since the blue point is contained in the black line.} 
    \label{fig:training_set_CDR}
\end{figure}

We show an example of a Clifford training set which provides a much more faithful mitigation with the ansatz including a constant term (see Fig.~\ref{fig:training_set_CDR}). Using an ansatz which contains the constant clearly allows for more flexible fitting of the training data. 

\section{Meson masses\label{sec:masses}}
In order to understand the phenomenon of confinement it is useful to project the Hamiltonian Eq.~\eqref{IisingHamiltonian} into the two kink subspace with basis $\{\ket{i,n}\}$:
\begin{align}
\mathcal{H}=&\sum_{i,n}\left[V(n)\ket{i,n}-Jh_X\left(\ket{i-1,n+1}\right.\right.\nonumber\\
&\phantom{\sum_{i,n}[}\left.\left.+\ket{i+1,n-1}+\ket{i,n-1}+\ket{i,n+1}\right)\right]\bra{i,n}
\label{eq:2-kink_H}
\end{align}
where $V(n)=2Jh_Zn$. The first term of this Hamiltonian represents a confining potential proportional to the separation between the domain walls and the second allows nearest neighbour interactions due to hopping. Therefore, a pair of kinks will experience an oscillatory motion due to the confining potential resulting in a meson.

In the case with $h_Z=0$, $\sigma^Z(t)$ decays to zero exponentially for any quench with $h_X<1$~\cite{calabrese_quantum_2011}. However, when a longitudinal field is introduced, the dynamics changes and an oscillatory behaviour is observed with various frequencies from which the masses of the two kinks bound states can be extracted. For this purpose, we consider the states $\ket{i,n}$ indicated in Eq.~\eqref{eq:meson} which are eigenstates of the Hamiltonian with $h_X=0$ and we perform a quench up to a certain value $h_X<1$. To obtain the eigenstates of the system after the quench we use the 2-kink model introduced before as it is a good approximation of the low energy behaviour of the system even for large values of $h_X$~\cite{tan2019observation}. To diagonalise the Hamiltonian Eq.~\eqref{eq:2-kink_H} we start by changing to the momentum space,
\begin{equation}
    \ket{k,n}=\frac{1}{\sqrt{L-(n+1)}}\sum_j^{L-(n+1)}e^{-ikj-ik\frac{n}{2}}\ket{j,n},
\end{equation}
so that the Hamiltonian becomes
\begin{align}
 \mathcal{H}=\sum_{k,n}&\left[V(n)\ket{k,n}\bra{k,n}+2h_X\cos\qty(\frac{k}{2})\right.\nonumber\\
 &\left.\left(\ket{k,n}\bra{k,n+1}+\ket{k,n}\bra{k,n-1}\right)\right].    
\end{align}
This Hamiltonian is diagonal in the basis of states
\begin{equation}
    \ket{k,\alpha}=\sum_n\mathcal{C}_\alpha\mathcal{J}_{n-\nu_{k,\alpha}}\qty(x_k)\ket{k,n}
\end{equation}
where $\mathcal{J}$ is the Bessel function of the first kind, $\nu_{k,\alpha}=E_{k,\alpha}/2h_X$, $x_k=2h_Z\cos\qty(\frac{k}{2})/h_x$ and $\mathcal{C}_\alpha$ is a coefficient to normalise the state~\cite{vovrosh_confinement_2020}. Therefore, $\ket{k,\alpha}$ are the eigenstates of the Hamiltonian with $h_X\neq0$ and $h_Z\neq0$ and we can write the state of the system at time $t$ as
\begin{equation}
    \ket{\Psi(t)}=\sum_{k,\alpha}\bra{k,\alpha}\ket{\Psi(0)}e^{-iE_{k,\alpha}t}\ket{k,\alpha}
\end{equation}
where $\ket{\Psi(0)}$ is the initial state. Using this expression, the expected value of a certain observable $\mathcal{O}$ is
\begin{align}
    \bra{\Psi(t)}\mathcal{O}\ket{\Psi(t)}=\sum_{\substack{k,\alpha\\q,\beta}}&\bra{\Psi(0)}\ket{k,\alpha}\bra{q,\beta}\ket{\Psi(0)}\cdot\nonumber\\
    &\cdot\bra{k,\alpha}\mathcal{O}\ket{q,\beta}e^{-i(E_{q,\beta}-E_{k,\alpha})t}
\label{eq:exp_val}
\end{align}
where we see an oscillatory behaviour with frequencies equal to energy differences between eigenstates. A method of obtaining the masses corresponding to different excited states, $m_\alpha=E_{0,\alpha}-E_{0,0}$, is to use initial states whose dominant oscillation frequency is $\omega_\alpha=E_{0,\alpha+1}-E_{0,\alpha}$ because then, the masses $m_\alpha$ are given by $m_\alpha=\sum_{\beta=0}^\alpha\omega_\beta$.

With the parameters $h_X$ and $h_Z$ that we are using, if we consider the initial states $\ket{i=4,n=1}=\ket{\uparrow\uparrow\uparrow\uparrow\downarrow\uparrow\uparrow\uparrow\uparrow}$ and $\ket{i=4,n=2}=\ket{\uparrow\uparrow\uparrow\uparrow\downarrow\downarrow\uparrow\uparrow\uparrow}$, the dominant frequencies are $\omega_1$ and $\omega_2$, respectively. The highest coefficients,
\begin{equation}
    c_{k,\alpha,n}=\frac{\mathcal{C}_\alpha\mathcal{J}_{n-\nu_{k,\alpha}}}{\sqrt{L-(n+1)}},
\end{equation}
in the expansion Eq.~\eqref{eq:exp_val} written in the basis $\ket{j,n}$ are those corresponding to the states $\ket{k=0,\alpha=2}$ and $\ket{k=0,\alpha=1}$ for the initial state $\ket{i=4,n=1}$ and $\ket{k=0,\alpha=3}$ and $\ket{k=0,\alpha=2}$ for the initial state $\ket{i=4,n=2}$.

It should be noted that we cannot use the two-kink approximation to obtain the energy of states with $n=0$, such as the state $\ket{\uparrow\uparrow\uparrow\uparrow\uparrow\uparrow\uparrow\uparrow\uparrow}$. However, a dominant oscillation frequency is also observed in the temporal evolution of the magnetisation using this initial state. This oscillation frequency is to be understood as $\omega_0=E_{0,1}-E_{0,0}$ since it corresponds to the energy required to create a particle with momentum zero~\cite{PhysRevA.95.023621}.

\begin{figure}[htb!]
    \centering
    \includegraphics[width = 0.99\linewidth]{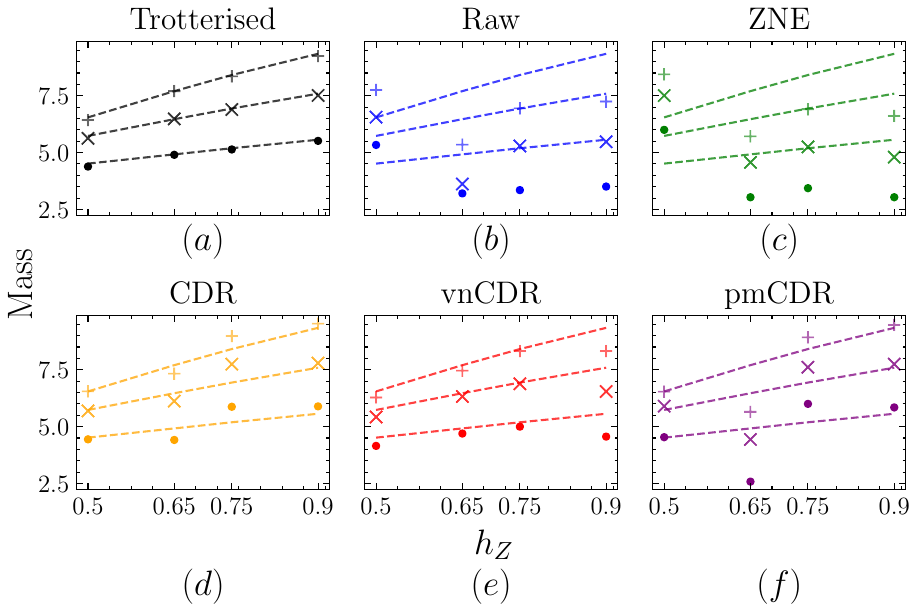}
    \caption{Masses obtained at $h_X=0.5$ and different longitudinal fields from the exact diagonalised (dashed lines), simulated (a), raw (b) and mitigated data (c)-(f). Masses obtained for initial states $\ket{\uparrow\uparrow\uparrow\uparrow\uparrow\uparrow\uparrow\uparrow\uparrow}$ (dots),  $\ket{\uparrow\uparrow\uparrow\uparrow\downarrow\uparrow\uparrow\uparrow\uparrow}$ (diagonal cross) and  $\ket{\uparrow\uparrow\uparrow\downarrow\downarrow\uparrow\uparrow\uparrow\uparrow}$ (vertical cross) are plotted.}
    \label{fig:mass_0.5}
\end{figure}

Following this prescription, we show in Fig.~\ref{fig:mass_0.5} the masses corresponding to the frequencies of Fig.~\ref{fig:freq_0.5}.

\section{Additional data \label{sec:data}}
\begin{figure}[htb!]
    \centering
    \includegraphics[width = 0.99\linewidth]{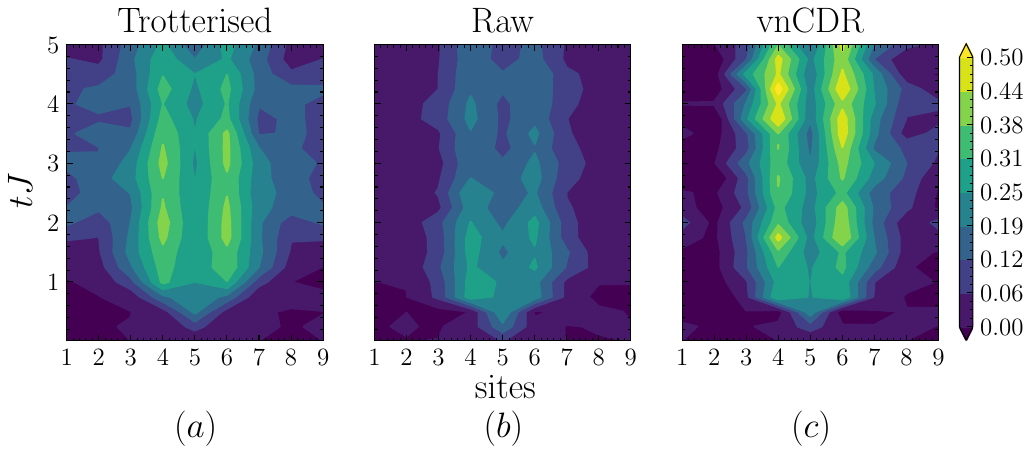}
    \caption{Correlation of qubits at sites along the x axis with the central qubit for the TFIM with $h_Z=0.2$ and the system initialised in the $\ket{\uparrow\uparrow\uparrow\uparrow\uparrow\uparrow\uparrow\uparrow\uparrow}$ state. The mean absolute error relative to the trotterised dynamics for the raw observables is $0.521$. vnCDR reduces this error to $0.402$. In this case vnCDR gave the best mitigated values for the correlation closely followed by CDR.}
    \label{fig:correlation_0.2}
\end{figure}

\begin{figure*}[htb!]
    \centering
    \subfloat{
        \includegraphics[width = 0.99\textwidth]{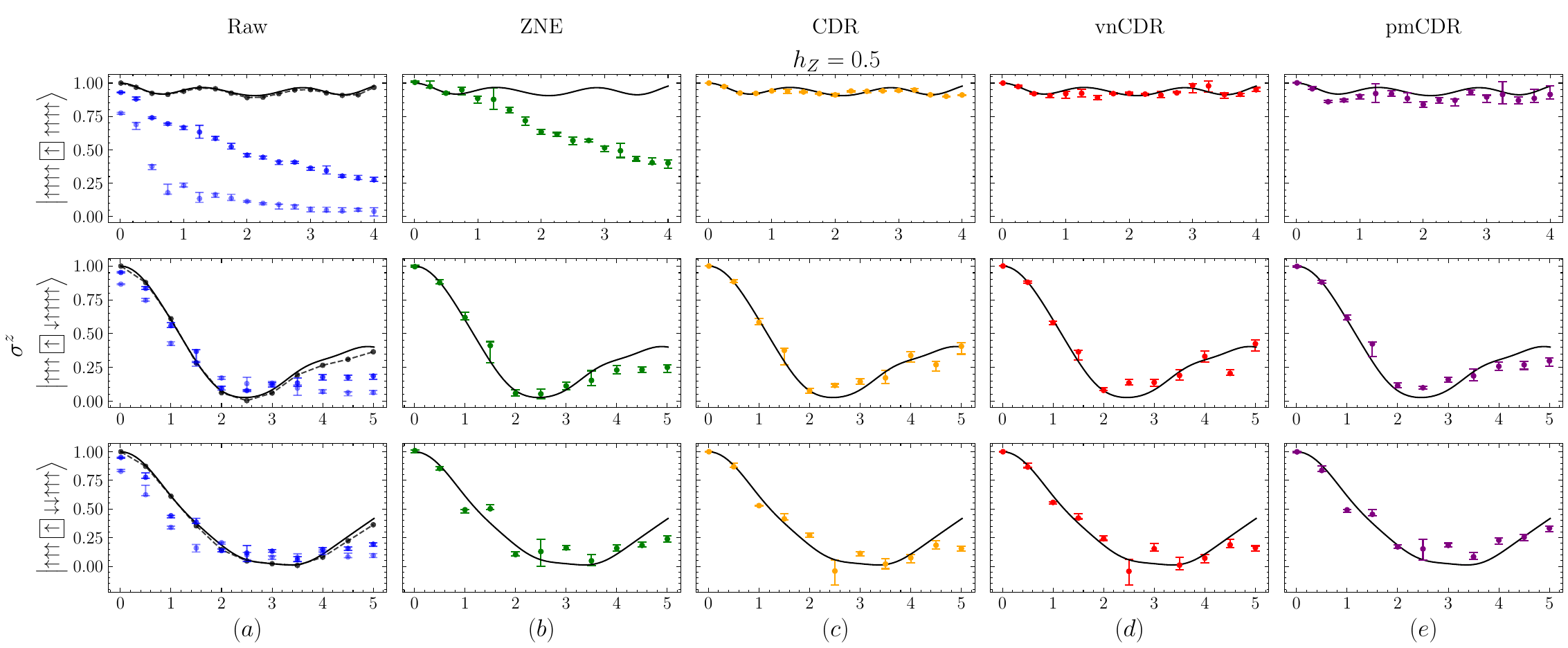}
    }\\
    \subfloat{
        \includegraphics[width = 0.99\textwidth]{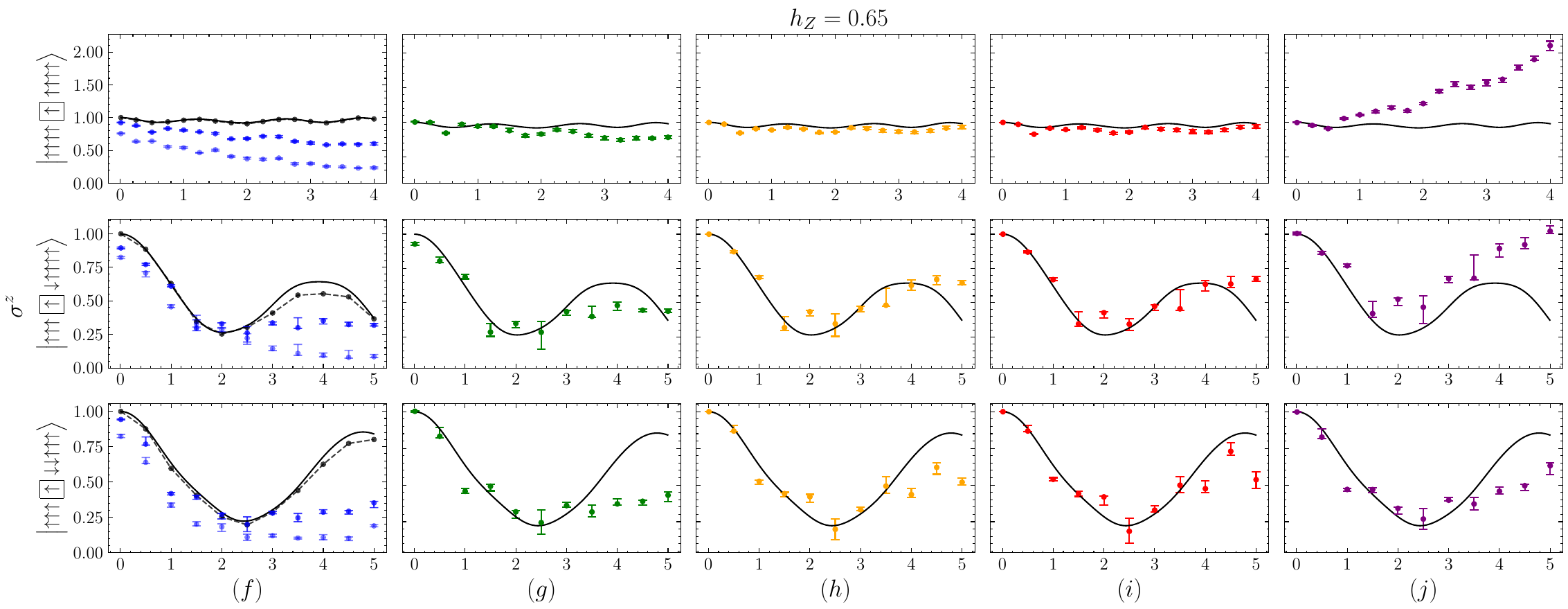}
    }\\
    \subfloat{
        \includegraphics[width = 0.99\textwidth]{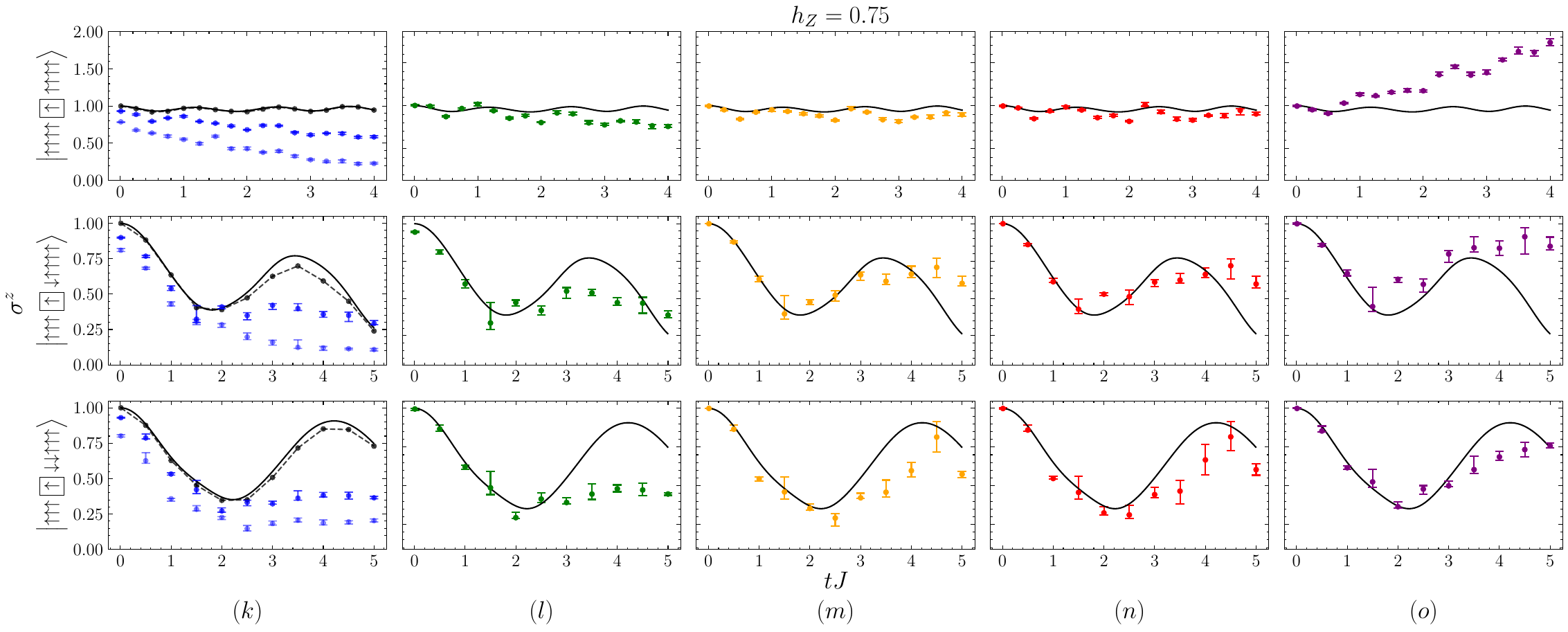}
        }
    \caption{Temporal evolution of $Z$-axis local magnetisation with $h_X=0.5$ and $h_Z=0.5,\text{ }0.65,\text{ }0.75$ for three initial states (in each row) with the observables mitigated using various techniques (columns). In all panels the exact dynamics is shown (black-solid line). Raw observables are shown in the left most column [($a$), ($f$), ($k$)] calculated at two noise levels $\mathcal{C} = \{1, 3\}$ (blue and light blue points respectively) using the Paris quantum computer. Black points and dashed lines show the exact trotterised dynamics. Error bars show the distribution of observables calculated over 6 repeats of the circuits of interest, and central points show the median. The raw observables at the first noise level were calculated to have a mean absolute error  of [($a$) $0.299$,  ($f$) $0.246$, ($k$) $0.248$] with respect to the trotterised dynamics, normalised by the average exact value. ZNE reduced this error to [($b$) $0.209$, ($g$) $0.163$ , ($f$) $0.148$], CDR to [($c$) $0.115$, ($h$) $0.123$, ($m$) $0.108$], vnCDR to [($d$) $0.117$, ($i$) $0.113$, ($n$) $0.106$] and pmCDR to [($e$) $0.134$, ($j$) $0.310$, ($o$) $0.241$].}
    \label{fig:magnetisation_0.5-0.65-0.9}
\end{figure*}

\begin{figure*}[htb!]
    \centering
    \includegraphics[width = 0.99\textwidth]{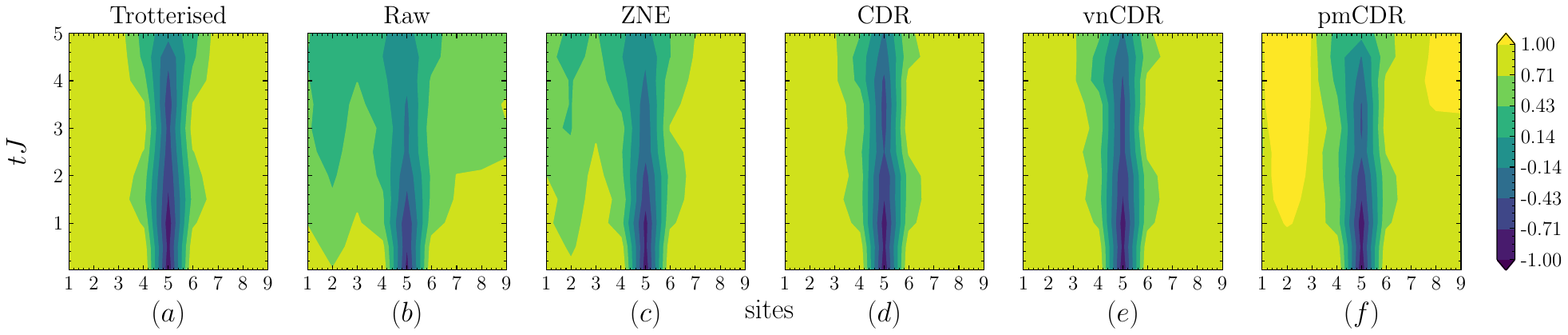}
    \caption{Temporal evolution of $Z$-axis local magnetisation with $h_X=0.5$ and $h_Z=0.9$ for the initial state $\ket{\uparrow\uparrow\uparrow\uparrow\downarrow\uparrow\uparrow\uparrow\uparrow}$ with the observables mitigated using various techniques (columns). The raw observable have a mean absolute error of $0.314$ with respect to the trotterised dynamics, normalised by the average exact value. ZNE reduced this error to $0.208$, CDR to $0.074$, vnCDR to $0.065$ and pmCDR to $0.235$.}
    \label{fig:magnetisation_0.9_Q9}
\end{figure*}

Here we present the results for the time evolution of the local magnetisation with different values of $h_Z$ used to obtain the frequencies shown in Fig.~\ref{fig:freq_0.5}. (see Fig.~\ref{fig:magnetisation_0.5-0.65-0.9}). We also show a colour plot of the magnetisation for each qubit in the system for $h_{X}=0.5, h_{Z} = 0.9$ (see Fig.~\ref{fig:magnetisation_0.9_Q9}). Furthermore, we show the additional results for the correlation when $h_{Z} = 0.2$ (see Fig.~\ref{fig:correlation_0.2}). The mitigation of the correlation here is a little worse than that presented in the main text. This could be attributed to the number of non-Clifford gates per Trotter step being greater when $h_{Z} \neq 0$. This means for the same circuit depth more gates need to be substituted when forming the training circuits, making for a less reliable dataset in general.
\section{Quantum circuit for trotterised evolution\label{sec:circuit}}
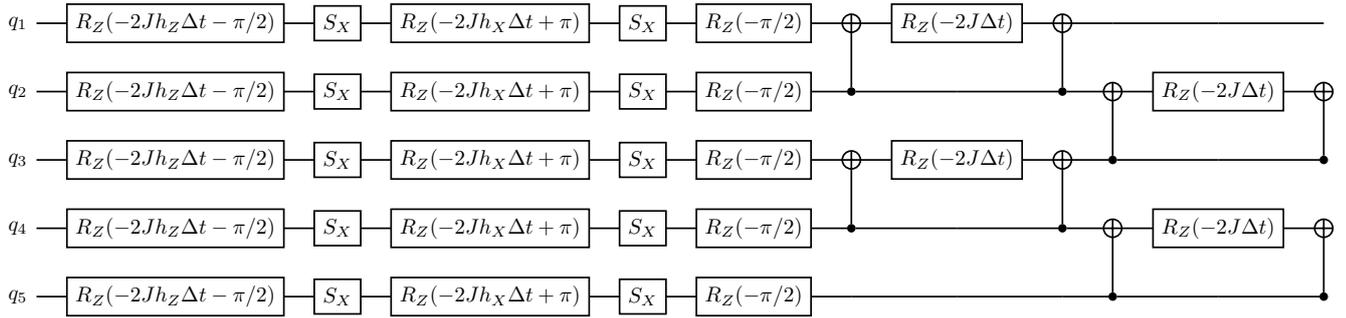
\begin{figure*}
\scalebox{.8}{
\begin{quantikz}
\lstick{$q_1$} & \gate{R_Z(-2Jh_Z\Delta t-\pi/2)} & \gate{S_X} & \gate{R_Z(-2Jh_X\Delta t+\pi)} & \gate{S_X} & \gate{R_Z(-\pi/2)} & \targ{} & \gate{R_Z(-2J\Delta t)} & \targ{} & \qw & \qw  & \qw \\
\lstick{$q_2$} & \gate{R_Z(-2Jh_Z\Delta t-\pi/2)} & \gate{S_X} & \gate{R_Z(-2Jh_X\Delta t+\pi)} & \gate{S_X} & \gate{R_Z(-\pi/2)} & \ctrl{-1} & \qw & \ctrl{-1} & \targ{} & \gate{R_Z(-2J\Delta t)} & \targ{}\\
\lstick{$q_3$} & \gate{R_Z(-2Jh_Z\Delta t-\pi/2)} & \gate{S_X} & \gate{R_Z(-2Jh_X\Delta t+\pi)} & \gate{S_X} & \gate{R_Z(-\pi/2)} & \targ{} & \gate{R_Z(-2J\Delta t)} & \targ{} & \ctrl{-1} & \qw  & \ctrl{-1}\\ 
\lstick{$q_4$} & \gate{R_Z(-2Jh_Z\Delta t-\pi/2)} & \gate{S_X} & \gate{R_Z(-2Jh_X\Delta t+\pi)} & \gate{S_X} & \gate{R_Z(-\pi/2)} & \ctrl{-1} & \qw & \ctrl{-1} & \targ{} & \gate{R_Z(-2J\Delta t)} & \targ{}\\
\lstick{$q_5$} & \gate{R_Z(-2Jh_Z\Delta t-\pi/2)} & \gate{S_X} & \gate{R_Z(-2Jh_X\Delta t+\pi)} & \gate{S_X} & \gate{R_Z(-\pi/2)} & \qw & \qw & \qw & \ctrl{-1} & \qw & \ctrl{-1}
\end{quantikz}}
\caption{ Quantum circuit representation showing one step of a first order Trotter expansion of a $5$-qubit encoded spin system.}
\label{fig:QCircuit}
\end{figure*}

The quantum circuit to evolve the system by one Trotter step for a $5$ spin system is shown in Fig.~\ref{fig:QCircuit}. In general for a system with an odd number of qubits $Q$ each Trotter step has a depth of $11$ with $2(Q-1)$ entangling gates  and $3Q-1$ non-Clifford gates.
\pagebreak
\section{Data collection}
\begin{figure*}[htb!]
    \centering
    \includegraphics[width = 0.99\textwidth]{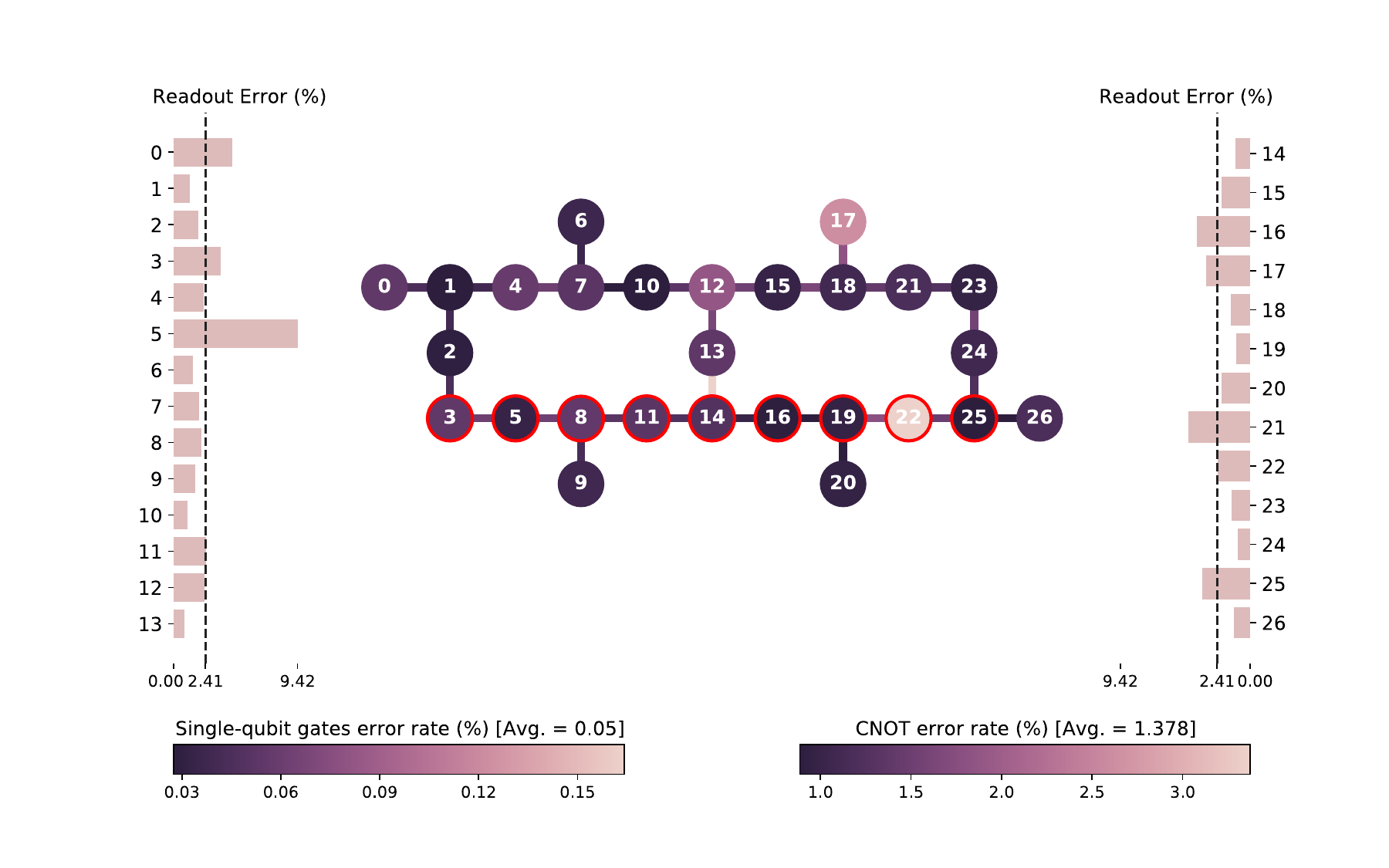}
    \caption{Errors on the IBMQ Paris quantum computer at the time the circuits in this work were run. The qubits we used to form a spin chain are encircled in red. For the qubits used in this work $T_{1} = 76 \pm 21 \mu s$ and $T_{2}= 65 \pm 26 \mu s$ were the mean parameters.}
    \label{fig:computer_errors}
\end{figure*}

We used the Paris quantum computers available through the IBM quantum cloud access. The data in this work was collected during the days between 22nd February 2021 and 5th March 2021 the errors in the Paris machine remained consistent over the times of collection and are shown in Fig.~\ref{fig:computer_errors}. We randomly selected the qubits to form a $9$ qubit chain. The results obtained here could be improved by choosing the qubits to use in a systematic manner, minimising the single qubit and CNOT error.


\end{document}